\documentclass{PoS}

\newcommand{\be}{\begin{equation}}
\newcommand{\ee}{\end{equation}}
\newcommand{\ba}{\begin{eqnarray}}
\newcommand{\ea}{\end{eqnarray}}
\newcommand{\bi}{\begin{itemize}}
\newcommand{\ei}{\end{itemize}}

\newcommand{\tr}{{\rm Tr\,}}
\newcommand{\re}{\mathop{\rm Re}}

\newcommand{\<}{\langle}
\renewcommand{\>}{\rangle}

\newcommand{\la}{\label}

\newcommand{\txts}{\textstyle}

\title{QCD at non-zero temperature from the lattice}

\ShortTitle{QCD at non-zero temperature from the lattice}

\author{\speaker{Harvey B.\ Meyer}\\
        PRISMA Cluster of Excellence, Institut f\"ur Kernphysik and Helmholtz Institut Mainz,
 Johannes~Gutenberg-Universit\"at~Mainz,  D-55099 Mainz, Germany \\ 
        E-mail: \email{meyerh@kph.uni-mainz.de}}

\abstract{I review the status of lattice QCD calculations at non-zero
  temperature.  After summarizing what is known about the equilibrium
  properties of strongly interacting matter, I discuss in more detail
  recent results concerning the quark-mass dependence of the thermal phase
  transition and the status of calculations of non-equilibrium properties.}

\FullConference{The 33rd International Symposium on Lattice Field Theory\\
		14 -18 July 2015\\
		Kobe International Conference Center, Kobe, Japan*}

\begin{document}

\section{Introduction\la{sec:intro}}
\vspace{-0.1cm}

In the 1960's, bulk hadronic matter was predicted to undergo a thermal
phase transition due to the high multiplicity of
hadrons~\cite{Hagedorn:1965st}.  With the advent of QCD as the theory
of the strong interaction, the phase above that transition
 was identified with a weakly interacting gas of
quarks and gluons~\cite{Cabibbo:1975ig} in virtue
of asymptotic freedom.  High-temperature QCD was
studied more systematically starting in the late
1970's~\cite{Linde:1980ts}.  The first studies of  
strongly interacting matter at non-zero temperature $T$ using lattice
QCD date back to
1981~\cite{Engels:1980ty}.  Lattice QCD is 
so far the only computational framework to have led to reliable
quantitative predictions on equilibrium quantities such as the
equation of state, the transition temperature $T_c$ and quark number
susceptibilities in the temperature range $100\lesssim T/{\rm
  MeV}\lesssim 500$.  Thereby it has had an important impact on
heavy-ion collision phenomenology.  Lattice QCD has also contributed
decisively to answering qualitative questions such as `What is the
nature of the thermal transition?', `What is the character of the
high-temperature phase at $T=250$MeV, 500MeV and 1000MeV?' or `What is
the character of the low-temperature phase at $T=140$MeV?'.  While it
is accepted that the medium is a gas of pions a sufficiently
low-temperatures, and a gas of quarks and gluons at sufficiently high
temperatures, lattice methods are the only theory methods able to
bridge the gap between these two limits.

I attempt to review the status of the lattice QCD effort. The
achievements are remarkable, yet there are still significant
challenges to be overcome before we can claim that all the qualitative questions
formulated above are answered in a satisfactory manner, particularly
concerning the non-equilibrium quantities.  A review of this length
cannot possibly be exhaustive. Fortunately, several reviews have
appeared in recent years~\cite{Brambilla:2014aaa,Meyer:2011gj,Soltz:2015ula,Ding:2015ona},
as well as lattice conference write-ups.


The next section gives an overview of what has been learnt about QCD
at non-zero temperature from the lattice; here the results are not
necessarily recent.  In section \ref{sec:colplot}, I cover recent work
on the quark-mass dependence of the phase diagram (Columbia
plot). Section \ref{sec:realtime} is devoted to non-equilibrium
quantities. The last section contains some final remarks.

\section{QCD at non-zero temperature: a portrait\la{sec:portrait}}
\vspace{-0.1cm}

We begin by fixing some notation. Lattice QCD employs the Matsubara
formalism of thermal field theory, where the partition function $Z
\equiv {\rm Tr}\{e^{-\beta \hat H}\}$ is represented as a Euclidean
path integral, in which bosons have periodic, fermions antiperiodic
boundary conditions in time~\cite{Kapusta:2006pm}.  The extent of the
time and space directions is $\beta=1/T=N_ta$ and $L$ respectively.
We will only consider isospin symmetric QCD, with up-down quark masses
$m_{ud}$ and strange quark mass $m_s$.  The isospin Pauli matrices are
denoted $\tau^a$ and $\psi$ is the up/down quark doublet field.  Let
$\<\bar\psi\psi\>$ be the chiral condensate.  In the chiral limit, it
is the order parameter for chiral symmetry breaking: it vanishes above
the chiral transition.  Its susceptibility can be written $\chi_m =
\frac{T}{L^3} \frac{\partial^2}{\partial m_{\rm ud}^2}\log Z$.
Thermodynamic quantities of interest include the energy density $e$,
pressure $p$ and entropy density $s$.  Quark number susceptibilities
can be written as $\chi_s = \beta \int d^3x \<J_0(x)J_0(0)\>$
($x_0\neq 0$), if $J_\mu$ is a conserved vector current. We will
address a few transport coefficients, such as the diffusion
coefficient $D$ and the shear viscosity $\eta$.

\subsection{Properties at physical $u,d,s$ quark masses}

\noindent\emph{The transition temperature.---}
The most thorough investigations of the transition temperature have  
been performed with staggered fermions.  
The BW collaboration~\cite{Borsanyi:2010bp} locates the
transition temperature at $155(2)(3)$MeV, based on the inflection
point of $m_{ud}(\<\bar\psi\psi\>|^T_0)/T^4$.  
As for results based on the peak of the chiral susceptibility, the
same publication~\cite{Borsanyi:2010bp} gives $T_c=147(2)(3)$MeV,
while the HotQCD collaboration~\cite{Bazavov:2011nk} obtained
$T_c=154(8)(1)$MeV.  In the latter calculation, an additional feature
was the use of the expected scaling laws near the chiral limit.

Last year, a calculation~\cite{Bhattacharya:2014ara} using domain-wall
fermions yielded the result $155(1)(8)$MeV based on the same quantity,
thus providing a universality test, even though this calculation was
performed at a single, fixed value of $N_t=8$.
Calculations with Wilson fermions are progressively moving 
toward physical quark masses~\cite{Borsanyi:2015waa}.

\noindent\emph{Deconfinement.---}
It is an intriguing question whether the deconfinement of partons 
occurs at the same temperature as the chiral transition.
Within QCD, the question cannot be given a completely sharp meaning, since 
the transition is a crossover~\cite{Aoki:2006we,Bhattacharya:2014ara}, and 
there is no known strict order parameter for confinement.
Nevertheless, the light-quark number susceptibility
$\chi_s^{u,d}/T^2$ admits an inflection point 
practically at the same temperature as the chiral restoration~\cite{Bazavov:2012jq,Borsanyi:2011sw}.
The rise of the strangeness fluctuations as a function of temperature 
is delayed by about 20MeV.
Ratios of cumulants of the baryon number $R^B_{42}=\chi_4^B/\chi_2^B$, which are identically unity in
the hadron-resonance gas model, provide another useful criterion.  The ratio $R_{42}^B$
deviates substantially from unity already at $T=155$MeV~\cite{Borsanyi:2013hza}.  

\noindent\emph{Thermodynamic potentials.---}
Good agreement was reached between the calculations of the BW~\cite{Borsanyi:2013bia} and the HotQCD~\cite{Bazavov:2014pvz}
collaborations once the continuum limit was taken based on the ranges $6\leq N_t\leq 16$ and $6\leq N_t\leq 12$ respectively.
At $T=255$MeV, that is, 100MeV above the chiral transition, the pressure amounts to
about half the Stefan-Boltzmann pressure corresponding to free quarks and gluons.
And, the ratio\footnote{This quantity corresponds to the ratio of the 
expectation value of the trace of the energy-momentum tensor and the 00-component
of its traceless part.} $ (e-3p)/[\frac{3}{4}(e+p)]$ evaluates to about a third.
There are thus large deviations from both a weakly-coupled regime and from 
a scale-invariant regime.

\noindent\emph{The hadron resonance gas (HRG) model vs.\ lattice data.---}
The HRG model assumes that the thermal medium consists of
quasiparticles with a spectrum identical to the hadron spectrum in
vacuum. It is justified to some extent by the studies showing that
resonances whose width is narrow compared to the temperature can be
treated as constituents of the
medium~\cite{Dashen:1974jw}.  The model describes the  
thermodynamic potentials and the quark number susceptibilities very
well up to $T=155$MeV.  The level of agreement is surprisingly good,
given that most of the $T=0$ resonances are not narrow compared to the
temperature and that thermal effects are bound to modify the spectrum
of thermal quasiparticles.

\noindent\emph{Static screening masses.---}
In the high-temperature phase, a number of static correlation lengths
(=inverse screening masses) have been computed. The light-quark
isovector masses are expected to approach $2\pi T$ at high
temperatures. This behavior is approximately observed in several 
cal\-cu\-la\-tions~\cite{Cheng:2010fe,Gupta:2013vha,Brandt:2014uda,Brandt:2013mba,Kaczmarek:2013kva}.
However, looking more closely, the weak-coupling correction is of order $g^2(T)$, with a positive
coefficient~\cite{Laine:2003bd}.  What several lattice calculations have
found~\cite{Cheng:2010fe,Gupta:2013vha,Brandt:2014uda,Brandt:2013mba,Kaczmarek:2013kva}
 is that the pseudoscalar screening mass lies well below $2\pi
T$ at least until $T=500$MeV; higher temperatures have not been
explored much. Also, the transverse vector screening mass is found
to be below $2\pi T$ at $T=255$MeV~\cite{Brandt:2013faa}.
Therefore, from the point of view where the $z$ direction is interpreted 
as a Euclidean time, quarks appears to be more strongly bound than the weak-coupling 
treatment predicts.
Finally, we remark that calculations in the low-temperature phase indicate that the
pseudoscalar correlation length starts to become shorter well below
$T_c$, see e.g.~\cite{Brandt:2014qqa}.

\begin{figure}[t]
\centerline{\includegraphics[width=0.5\textwidth]{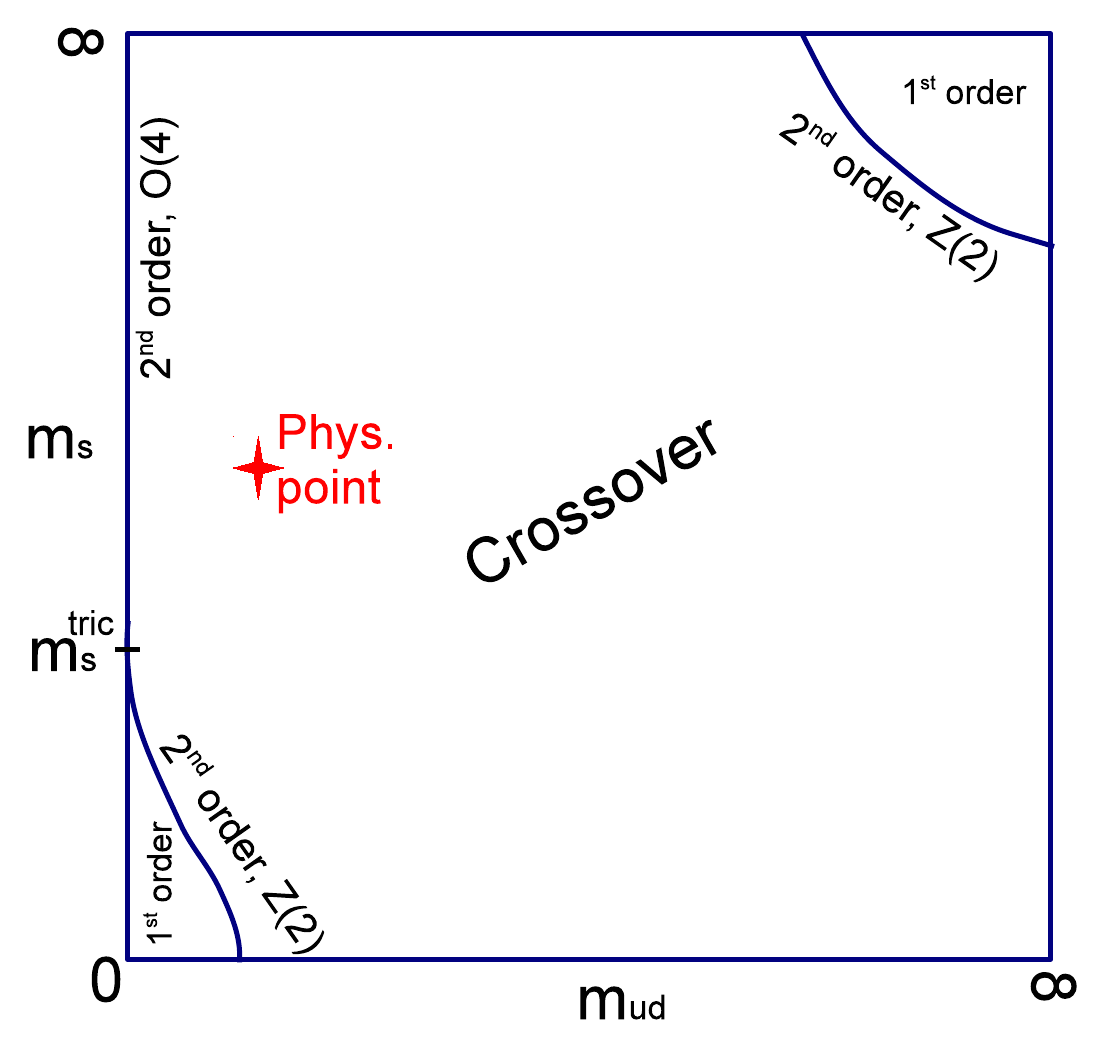}
\includegraphics[width=0.5\textwidth]{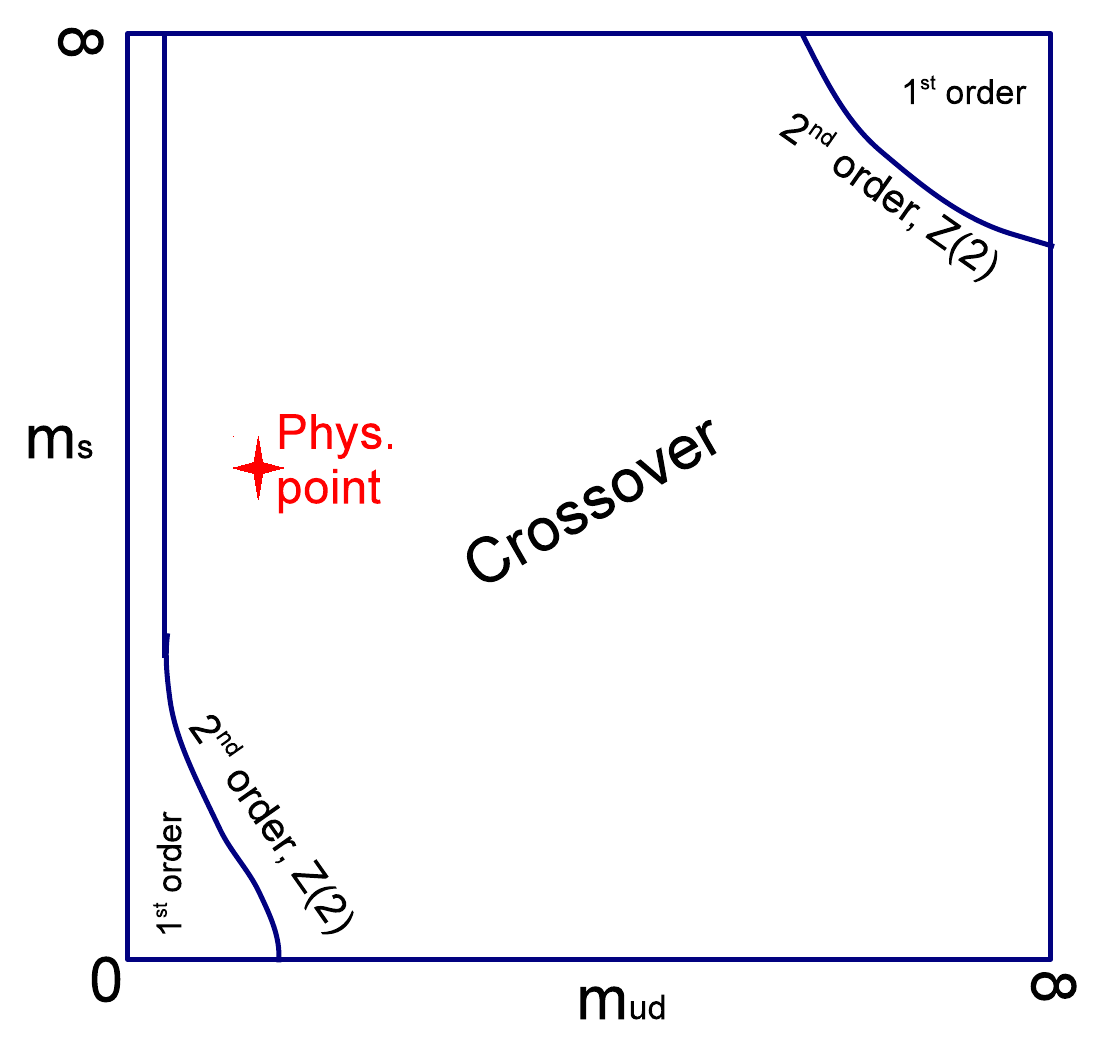}}
\caption{The `Columbia plot', which indicates the dependence 
of the nature of the thermal QCD transition on the up, down
and strange quark masses $m_{ud}\equiv m_u=m_d$ and $m_s$.
Two scenarios are depicted, which differ qualitatively in the $m_{ud}\to0$ limit. In the left, 
the phase transition is of second order at $m_{ud}=0$. In the right panel, the phase
transition is of first order below a critical $m_{ud}$ mass.}
\label{fig:columbia}
\end{figure}

\subsection{Dependence of the phase transition on the matter content of the gauge theory}

The simplest deformation of QCD with $u,d,s$ quarks consists in
varying the masses of these quarks. The nature of the thermal
transition is expected to have a specific dependence on these masses,
which is represented graphically in the `Columbia plot'; see
Fig.\ \ref{fig:columbia}.

Two features are firmly established, both by universality arguments
and by direct Monte-Carlo simulations. In the limit of infinite quark
masses, the theory reduces to the SU(3) pure gauge theory, and the
Z(3) symmetry associated with the center of the gauge group becomes
exact. As anticipated~\cite{Yaffe:1982qf}, based on the fact that Z(3)
invariant theories in three dimensions have first order phase
transitions, the deconfinement transition is found to be first order.
At the opposite corner of the Columbia plot, i.e.\ in the limit where
$m_{ud}\to0$ and $m_s\to 0$, a sharp phase transition, now associated
with chiral symmetry, takes place. There are strong arguments that it is 
first order~\cite{Pisarski:1983ms}.

In a broad range of $m_{ud}$ and $m_s$, including the physical values
of the masses, the thermal transition is a crossover. The Z(2)
critical line at large quark masses, where the first order transition
turns into a crossover, has been studied on the lattice. Pinning down
the Z(2) critical line in the lower left corner of the diagram is the
subject of current simulations; see section \ref{sec:colplot}.

In the SU(2) chiral limit, i.e.\ when $m_{ud}\to0$ with the strange
quark mass held fixed at its physical value, the chiral transition
must be a well-defined phase transition.  The transition may be of
second order in the O(4) universality class, or it may be of first
order~\cite{Pisarski:1983ms}. Which scenario is realized in
Fig.\ \ref{fig:columbia} is also the subject of current calculations;
see section \ref{sec:colplot}.

\subsection{Transition temperature as a function of the matter content}

The crossover temperature at physical $u,d,s$ quark masses $T_c\approx 155$MeV is 
lower than educated guesses made in the 1980's~\cite{Gasser:1986vb}. 
In the following we give an overview of the dependence of the transition temperature
on the matter content. For this purpose, we compare the $T_c$'s of different theories
in units of the Sommer reference scale $r_0$.
However, since MeV units are more familiar, we set $r_0\doteq 0.5$fm to quote 
results in MeV.

Within the three-flavor theory, extensive
calculations~\cite{Borsanyi:2015waa} with Wilson fermions show that
the transition temperature decreases as a function of the light-quark
masses.
With two flavors, results have been obtained recently with Wilson
fermions, both with and without a twisted-mass action. At a quark mass
corresponding to a zero-temperature pion mass of about 300MeV, the
crossover is very broad.  Ref.\ \cite{Brandt:2013mba} quotes
$T_c=211(5)$MeV, while the result of Ref.\ \cite{Burger:2014xga}
corresponds to $T_c=180(12)$MeV in our chosen scale setting.  There is
therefore some tension between these two results. The first was
obtained at finite $N_t=16$, the second is a continuum-extrapolated
result based on $N_t\leq 12$.

When all three quark masses become heavy, one reaches the SU(3) pure
gauge theory.  A recent, accurate result for the deconfinement
temperature is $T_c=294(2)$MeV~\cite{Francis:2015lha}.
In view of these results, one may wonder if the transition temperature
would decrease even further if light matter fields were added. This
question has arisen in the context of walking gauge theories. It has
been conjectured that the transition temperature goes all the way down
to zero as the number of flavors is increased; see the results and the
discussion in~\cite{Lombardo:2015oha}. Once $N_{\rm f}$ is large enough for
the transition to occur at $T=0$, the infrared properties of the $T=0$
system are described by a conformally invariant theory. The
phase transition has been used as a tool to study gauge theories with
many matter degrees of freedom, since it is not easy to numerically
distinguish a chirally broken phase from the conformally invariant
phase; see for instance~\cite{Kogut:2015zta}.


\subsection{Gauge group dependence of the equilibrium properties}

Another direction in which one can depart from QCD is by changing the gauge group.  
Varying the number of colors $N_c$, several studies of the
thermodynamic potentials of SU($N_c$) gauge theory have been
performed~\cite{Bringoltz:2005rr}.  
In the high-temperature phase, it is useful to define the function
\be
\bar p(x) = \Big(\frac{p}{p_{\rm SB}}\Big)(T=xT_c), 
\ee
whose value is the pressure normalized by the Stefan-Boltzmann pressure, as a function 
of the temperature in units of the transition temperature.
The result is that within the small uncertainties, $\bar p$ for $3\leq N_c\leq 8$  
is found to be `universal' for $x>1.1$. This empirical fact indicates that the multiplicity
of the degrees of freedom scales with $(N_c^2-1)$, as expected in the weak-coupling regime.
The departure from unity is still large, for instance $\bar p(1.6) \approx 0.5$, a value 
very similar to QCD with physical $u,d,s$ quark masses. 

Studies have also been performed with other gauge groups, most notably
with the exceptional group
G(2)~\cite{Pepe:2006er,Bruno:2014rxa}.  The theory admits  
color singlet asymptotic states, but has no center symmetry.
Remarkably, here too $\bar p(x)$ for $x\geq 1.1$ is consistent with
the function obtained for the SU($N_c$) groups~\cite{Bruno:2014rxa}.

In the low-temperature phase, it is interesting to ask whether a
`hadron' resonance gas model accounts for the equilibrium properties
as well as for realistic QCD.  According to the large-$N_c$ counting
rules, interactions among the `hadrons' should be of order $1/N_c^2$
in the pure SU($N_c$) gauge theories, and therefore even more
suppressed than in QCD, thus giving the model a stronger
justification.  While the low-lying glueball spectrum of SU(3) gauge
theory is fairly well known from lattice
calculations~\cite{Meyer:2008tr}, it turns 
out that, within the model, the thermodynamic potentials receive a
substantial contribution from very massive states.  Assuming that the
glueball spectrum extends at higher masses as predicted by the
closed-string spectrum, a good description of the thermodynamic
potentials computed on the lattice is
obtained~\cite{Caselle:2015tza,Meyer:2009tq}. Since the spectrum is
exponential (Hagedorn-type spectrum), the model prediction depend
somewhat on the value of the exponent chosen and on whether the
spectrum is assumed to be charge-conjugation symmetric.  Nevertheless
the agreement with lattice data is quite impressive.

\subsection{Very high temperatures: making contact with perturbation theory}

Since there are large deviations from the non-interacting behavior at
$T=1.6T_c$, one may wonder at what temperature contact is made with
perturbative calculations, which for many quantities have been pushed
to high orders.  In computing the pressure (or the quantity $e-3p$), a
multi-scale problem arises at very high temperatures, since a direct
approach would imply subtracting the vacuum contribution
non-perturbatively at the same lattice spacing. One
solution~\cite{Borsanyi:2012ve} consists in performing the vacuum
subtraction in several steps, adding and subtracting the pressure at
temperatures differing by a factor two.  Another approach consists in
computing the enthalpy $e+p$, which does not require a vacuum
subtraction. Then a renormalization factor is required, which may for instance 
be computed using shifted boundary
conditions~\cite{Giusti:2012yj,Robaina:2013zmb,Giusti:2014ila,Pepe:2015kda}. 
Overall perturbation theory provides a good description of the lattice
data for $T>5T_c$ if a coefficient of the O($g^6$) contribution is
fitted to the highest temperatures. However, the successive terms of
the perturbative series are alternating and hardly decreasing in magnitude with the
order -- this has been known for a long time~\cite{Kajantie:2002wa}.

Is the lack of apparent convergence of the perturbative series at
$T<10T_c$ a symptom that the nature of the medium is completely
different from the weak-coupling quark-gluon plasma picture? Probably
not. After all we know that the expansion is in powers of $g$, not
$\alpha_s=\frac{g^2}{4\pi}$, and that the medium has a
non-perturbative sector even at arbitrarily high temperatures.
Different observables are sensitive to this sector starting at
different orders.

\subsection{Summary}
\noindent $\bullet$ In the low-temperature phase, the hadron resonance gas model provides a good description 
of the equilibrium properties.
\\ $\bullet$  The thermal transition is a crossover.
\\ $\bullet$ The chiral and deconfinement transitions essentially coincide, with $T_c=155$MeV being a typical value 
obtained for the crossover.
\\ $\bullet$ Above the transition, the multiplicity of the degrees of freedom is consistent with the number
  expected for quarks and gluons;
\\ $\bullet$ but many equilibrium quantities are far from the weak-coupling predictions at least until $T=2.5T_c$.
In addition, the size of the elliptic flow observed in heavy-ion
collisions points to a medium with a very small shear viscosity to
entropy density ratio in the range $T_c<T<2.5T_c$, typical estimates being
$\eta/s\approx 0.12$ at RHIC and $\eta/s\approx 0.2$ in the ALICE experiment~\cite{Gale:2013da}. 
Taken together, the last two observations indicate that,
in the range of temperatures explored in heavy-ion collisions,
the partonic degrees of freedom are strongly correlated.
\vspace{-0.3cm}

\section{Toward a quantitative version of the Columbia plot\la{sec:colplot}}

\noindent$\bullet~$\emph{The region around the SU(3) chiral limit $m_{ud}=m_s=0$.---}
Lattice calculations find that for sufficiently small $u,d,s$ quark
masses the transition is first order.  However, quantitatively the
findings differ wildly depending on the lattice action and lattice
spacing used. Generally, the observation has been that a smaller
lattice spacing leads to a smaller first-order region.

Jin et al.\ have published~\cite{Jin:2014hea} and reported at this
conference the result of a finite-size scaling study at $N_t=6$ and 8,
including a continuum extrapolation.  They find $m_\pi^{\rm crit}
= 304(7)(14)(7)$MeV for the critical pseudoscalar mass.  This result
was obtained with an O($a$) improved Wilson fermion action and the Iwasaki
gauge action.
H.-T.\ Ding for the Bielefeld-BNL-CCNU collaboration, on the other hand, has reported $m_{\rm
  ps}^{\rm crit} \lesssim 50$MeV with the HISQ action at $N_t=6$,
based on infinite-volume scaling with Z(2) exponents. The data extends
impressively to $m_\pi = 80$MeV.

In view of the contradiction between these results, there is a real
need for simulations on finer lattices, $N_t\geq 12$, including a
continuum extrapolation, to quantify the size of the first-order
region.

\noindent$\bullet~$\emph{The SU(2) chiral limit at fixed, physical $m_s$.---}
The HotQCD collaboration~\cite{Ding:2013lfa,Ding:2013nha} reports on a
calculation with the HISQ action at $N_t=6$, in which the chiral
susceptibility shows no sign of a first-order phase transition, even
at $m_\pi=80$MeV. The results thus point to the standard scenario.
The observables are fitted with O(2) exponents rather than O(4) exponents, 
due to the taste-splitting effects of the staggered action.

\noindent$\bullet~$\emph{The chiral limit in the $N_{\rm f}=2$ theory.---}
In the limit $m_s\to \infty$ one reaches the two-flavor theory.  Here
there have been several calculations investigating the nature of the
thermal phase transition in a regime as chiral as possible.  O($a$)
improved Wilson fermions at $N_t=16$ lead to a crossover transition
for $m_\pi = 295$MeV~\cite{Brandt:2013mba}, and so do twisted mass
Wilson fermions for a pion mass of $333$MeV~\cite{Burger:2014xga}.
Using imaginary chemical potential and the predictivity of tricritical
scaling, \mbox{Bonati} et al.\ found~\cite{Bonati:2014kpa} a critical
pion mass of more than 600MeV with Wilson fermions at $N_t=4$.
Obviously, finite-lattice-spacing effects are a major issue here too.
Ch.\ Pinke and Ch.\ Czaban have reported on progress in these studies at $N_t=4$ and $N_t=6$
at this conference.

The effects of the anomalous breaking of the $U_A(1)$ symmetry are
expected to be gradually reduced as the temperature increases.  A light
$\eta'$ screening mass provides a mechanism that could make the
transition first order when $m_{ud}\to0$.  In my view, while it is
certainly interesting to study measures of the anomalous symmetry
breaking at $T\neq 0$, discriminating between the two scenarios of 
Fig.\ \ref{fig:columbia} still requires direct simulations at small $m_{ud}$.

\vspace{-0.3cm}

\section{Near-equilibrium properties\la{sec:realtime}}
\vspace{-0.1cm}

 While there is hardly an alternative to lattice QCD when it comes to
first-principles predictions for equilibrium properties,
out-of-equilibrium aspects of QCD at non-zero temperature are only
indirectly accessible on the lattice.
Let $J(x)$ be a local bosonic Hermitian operator; the 
relation between the Euclidean correlator $G$
and the spectral function $\rho$ reads
\be\la{eq:chsh}
G(x_0,{\vec p}) = \int d^3x~ e^{-i {\vec p\cdot\vec x}}\Big\langle J(x)J(0)\Big\rangle {=} 
\int_0^\infty {d\omega}\; \rho(\omega, {\vec p})\;K(x_0,\omega),
\quad K(x_0,\omega) = \frac{\cosh[\omega(\beta/2-x_0)]}{2\pi\;\sinh[\omega\beta/2]}.
\ee
This equation represents an inverse problem for $\rho(\omega,\vec p)$, given the correlator $G$.
In thermal equilibrium, its (Kubo-Martin-Schwinger, KMS) relation to the Wightman correlator
\be\la{eq:KMS}
\rho(\omega,{\vec p})  = (1-e^{-\beta \omega})\int_{-\infty}^\infty {dt} \int d^3x 
\;e^{i\omega t-i{\vec p}\cdot {\vec x}}\;\frac{1}{Z}\tr\{e^{-\beta \hat H}\;\hat J(t,{\vec x})\;\hat J(0)\}
\ee
shows explicitly how it encodes real-time information\footnote{In Eq.\ (\ref{eq:KMS}),
real-time evolution is implied, $ \hat J(t) = e^{i\hat Ht} \hat J\, e^{-i\hat Ht}$.
The hadronic tensor $W^{\mu\nu}(q,p) = \frac{1}{2}\sum_{\sigma_N}\int d^4x \; e^{iq\cdot x}\;\<N|\hat
J^\mu(x)\,\hat J^\nu(0)|N\>$, which determines the reaction $\gamma^* \,{\rm Nucleon}
\to X$, is a similar object.}. 
Inserting a complete set of (unit-normalized) energy eigenstates $|n\>$, 
the formal expression for the spectral function reads
\be\la{eq:RhoSpecRep}
\rho(\omega,\vec p) = \frac{4\pi}{L^3\,Z} \sinh(\beta\omega/2)\sum_{n,m} 
\Big|\Big\<n\Big|\int_{\vec x} e^{-i\vec p\cdot\vec x}\hat J(\vec x)\Big|m\Big\>\Big|^2\;
e^{-\beta(E_n+E_m)/2}\delta(\omega-(E_n-E_m)).
\ee
Such expressions are useful to prove formal relations, such as the
connection with the Minkowski-space correlators.  However, what we are
after are the collective excitations of the medium, which are in general not related 
in a simple manner to the states $|n\>$. The collective excitations are typically
poles in the frequency variable of the infinite-volume correlator, which depend on
the temperature and are not related in a simple way to the $|n\>$. For instance,
the hydrodynamic excitations associated with conserved currents have to be there.
Furthermore, quasiparticles may exist in the medium: there may be a pole at $(\omega=\omega_{\vec p})$ 
with ${\rm Im} (\omega_{\vec p})  \lesssim {\rm Re}(\omega_{\vec p})$; in that case,
$v_{\rm g} = \frac{d\omega_{\vec p}}{d|\vec p|}$ is its group velocity. 
There are media, however, with no quasiparticles.

\begin{figure}[t]
\vspace{-0.5cm}
\centerline{ \begin{minipage}{0.61\textwidth} \includegraphics[width=\textwidth]{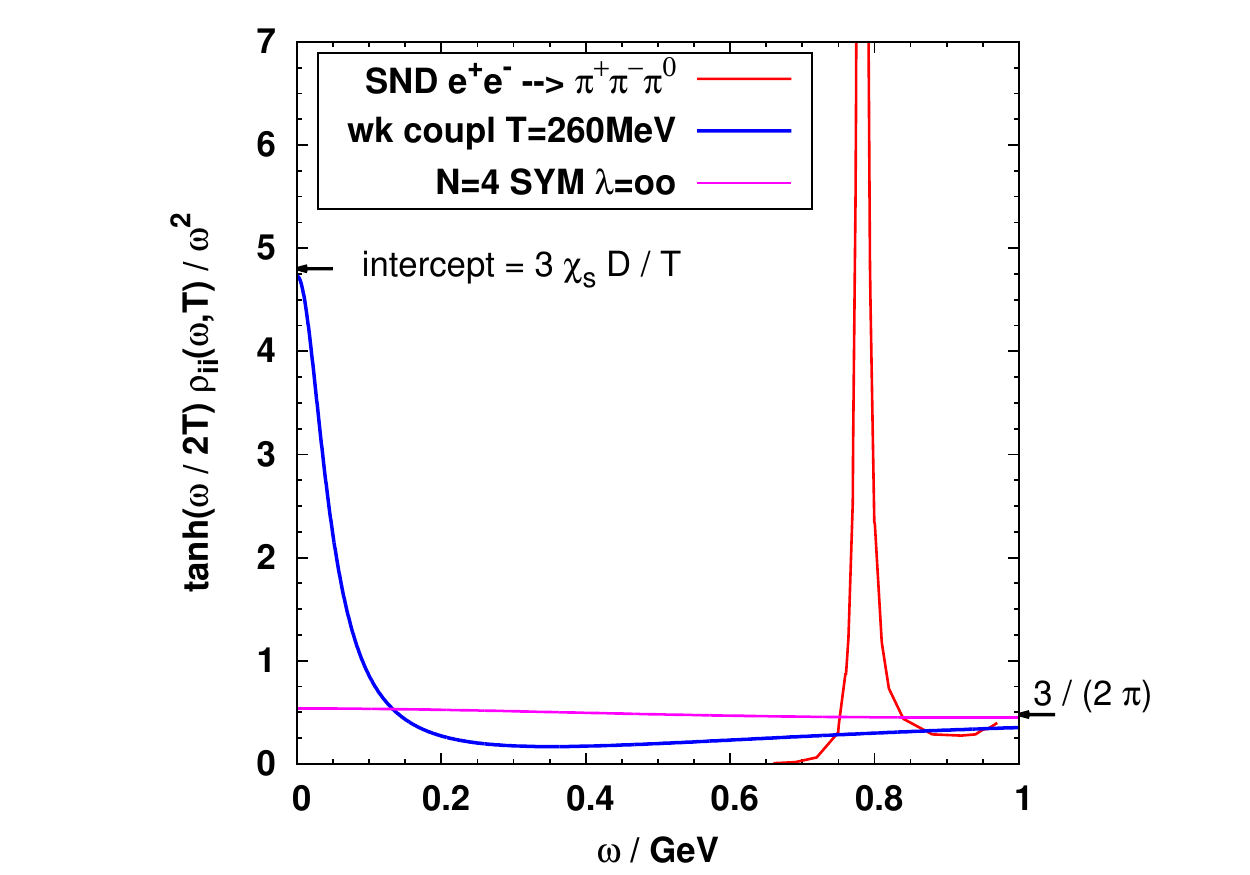}
\end{minipage}
\hspace{-1.2cm}
\begin{minipage}{0.66\textwidth}\vspace{0.3cm}\includegraphics[width=\textwidth]{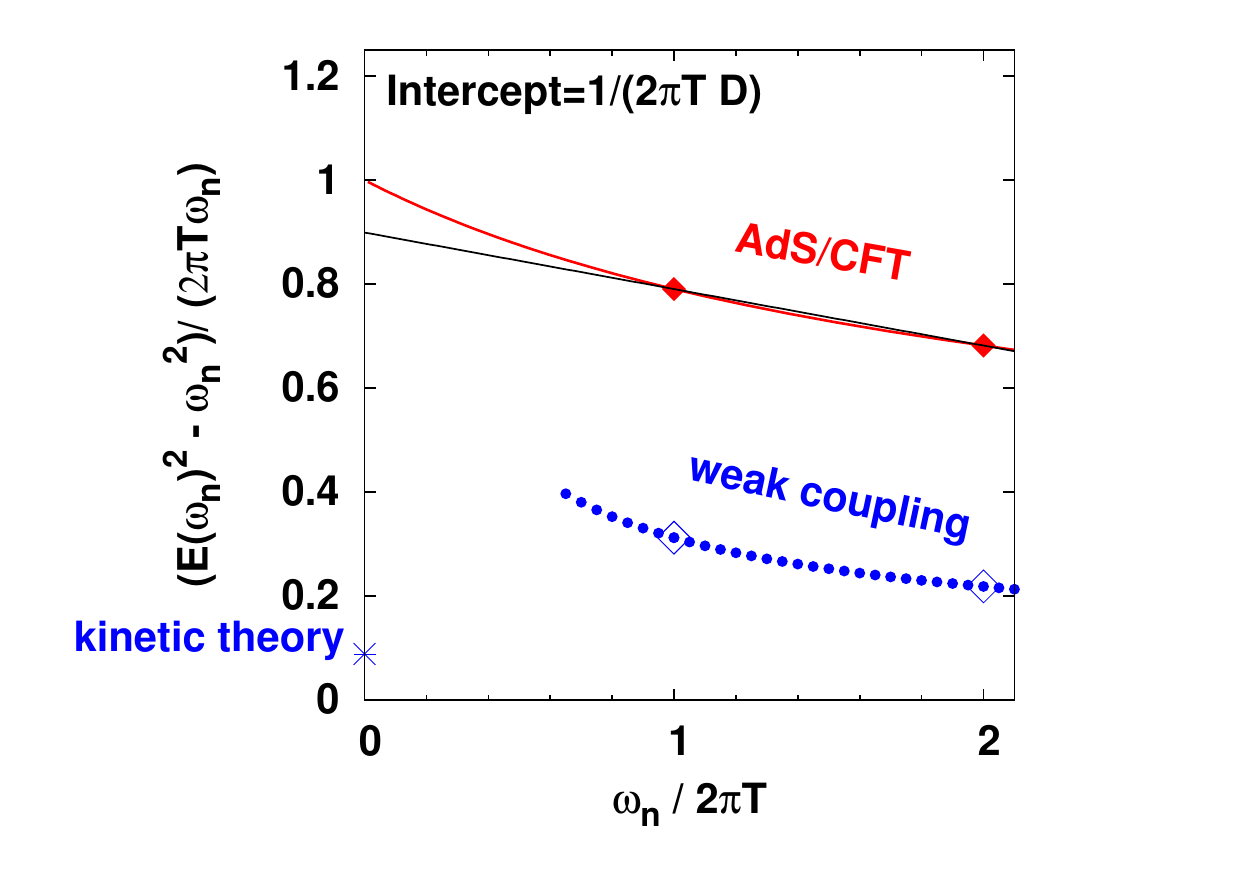}\end{minipage}}
\vspace{-0.2cm}
\caption{{\bf Left}: The spectral function of the current $J_i=\frac{1}{\sqrt{2}}(\bar u\gamma_i \bar u + \bar d \gamma_i d )$
at zero temperature ($e^+e^-$ data from~\cite{Achasov:2003ir}) 
and in the high-temperature phase of QCD at $T=254$MeV. In the latter case two `scenarios'
are depicted: one is the spectral function of the $R$-charge current~\cite{Myers:2007we} in 
the large-$N_c$ ${\cal N}=4$ super-Yang-Mills theory at infinite coupling (multiplied by the factor $(3/N_c)^2$);
the other is the QCD weak-coupling prediction, which contains a transport peak~\cite{Moore:2006qn} 
due to the existence of quark quasiparticles, and whose large-$\omega$ asymptotic value is $3/(2\pi)$.
Which scenario is closer to reality at temperatures accessed in heavy-ion collisions is 
of central importance to characterize the quark-gluon plasma. {\bf Right}:
{The screening mass $E(\omega_n)$ in non-static Matsubara
sectors $\omega_n\neq 0$. The blue curve corresponds to QCD at weak-coupling
at a temperature of 254MeV~\cite{Brandt:2014uda}. The red curve corresponds to the strongly coupled ${\cal N}=4$
SYM theory~\cite{Kovtun:2005ev,Brandt:2014cka}. The continuation of $E(\omega_n)$ down to $\omega_n=0$ 
is expected to yield the inverse diffusion coefficient: 
the kinetic-theory prediction of \cite{Arnold:2001ms} is displayed for the weak-coupling case.
The black (straight) line going through the first two non-static screening masses directly accessible
in the Matsubara formalism shows that a linear extrapolation provides a decent approximation in the strongly
coupled case. Both the blue and the red curve tend to 0 for $\omega_n\to\infty$.
}
}
\label{fig:isosv}
\end{figure}

The inverse problem is numerically ill-posed. The success of the
calculation thus depends on what resolution in the frequency $\omega$
one needs to achieve. Physically, resolving the spectral function on a
finer frequency scale means that one understands the real-time evolution of a particular
excitation up to a longer time-scale.  As an illustrative example, we consider the isoscalar
vector channel with the current $J_i=\frac{1}{\sqrt{2}}(\bar u\gamma_i
\bar u + \bar d \gamma_i d )$.  At $T=0$, detailed information on the
spectral function is provided by $e^+e^-$ collider data.  For
$\omega<1{\rm GeV}$, the spectral function is completely dominated by
the $\omega$ resonance; see Fig.\ \ref{fig:isosv}, left panel. The spectral
function vanishes below the threshold $\omega=3m_\pi$, and remains
very small at least until $\omega=700$MeV. These features are
characteristic of QCD at low energies, and reflect the confinement of
quarks. Correspondingly, one expects a very different form of the
spectral function in the high-temperature phase. At sufficiently high
temperatures, quarks and gluons are expected to be good
quasiparticles. They lead to a characteristic `transport
peak'~\cite{Aarts:2002cc,Moore:2006qn} around $\omega=0$ in the
spectral function of some channels; on the other hand, no meson-like
excitations are expected, so that the spectral function goes over
smoothly to the high-frequency perturbative behavior.  The expected
behavior is represented by the blue curve in Fig.\ \ref{fig:isosv}.
However, there are strongly-coupled gauge theories where, thanks to
the AdS/CFT correspondence, the spectral function can be computed
(semi-)analytically~\cite{Myers:2007we}. The magenta curve on the figure shows that no
transport peak is present and the function
$\tanh(\beta\omega/2)\rho_{ii}(\omega)/\omega^2$ is almost
constant. This behavior reflects the absence of
quasiparticles. Whether the QCD spectral function at temperatures
$200\lesssim T/{\rm MeV}\lesssim 500$ accessed in heavy-ion collisions
already admits a transport peak, or instead resembles more a
quasiparticle-free spectral function such as the magenta curve is
one of the central questions to answer in thermal QCD.

As a related question, one would like to know from what temperature on
the weak-coupling shape of the spectral function sets in.  It is also
not a priori clear at what temperature the $\omega$ and other narrow mesons
disappear; they may already have disappeared at $T=155$MeV, or instead
might survive for a short interval of temperature. The
transition to a medium without light meson-like excitations can also
be taken as a definition of deconfinement. For bound states of heavy
quarks ($\bar b b$), due to their small size the dissociation
temperature is parametrically higher.

\noindent\emph{Aspects of the inverse problem.---}
In practice the correlator is known at some discrete points in time, $G(x_0^{(i)})$.
Exploiting the linearity of the problem, we may take linear combinations of Eq.\ (\ref{eq:chsh}),
\be\la{eq:lincomb}
\sum_{i=1}^n c_i(\bar\omega)\, G(x_0^{(i)}) = \int_0^\infty d\omega\, \rho(\omega)\, 
{\widehat\delta(\bar\omega,\omega)} \equiv \hat\rho(\bar\omega),
\qquad 
\widehat\delta(\bar\omega,\omega)={\sum_{i=1}^n} c_i(\bar\omega)\,K(x_0^{(i)},\omega).
\ee
One may now choose the coefficients $c_i(\bar\omega)$ so that the
`resolution function' $\widehat\delta(\bar\omega,\omega)$ is as
narrowly peaked around a given frequency $\bar\omega$ as
possible. This is the idea behind the Backus-Gilbert method, which was
recently used in the lattice QCD context~\cite{Brandt:2015sxa} (see
Refs.\ therein).  One obtains a `locally smeared' version
$\hat\rho(\bar\omega)$ of the spectral function. The optimal
resolution function only depends on the kernel and the $x_0^{(i)}$;
the required coefficients, however, are then wildly oscillating,
requiring a regularization. An important feature of the method is that
it makes no assumption about the functional form of the spectral
function. In fact, in any finite volume the spectral function is a
distribution (see Eq.\ (\ref{eq:RhoSpecRep})), and only
$\hat\rho(\bar\omega)$ has a smooth infinite-volume
limit~\cite{Meyer:2011gj}. Note that it is in general not true that
integrating $\hat\rho(\omega)$ with the kernel $K(x_0,\omega)$ yields
a good description of the lattice correlator. It is thus 
also desirable to produce physically reasonable instances of
spectral functions which describe the data with a good $\chi^2$.

Usually it is assumed that the spectral function can be approximated
by a smooth function, e.g.\ via an explicit ansatz or the maximum
entropy method~\cite{Asakawa:2000tr}.  A new Bayesian method was
proposed in~\cite{Burnier:2013nla}; see the talk by S.\ Kim at this
conference. Stochastic methods are being imported from other fields
and adapted to lattice QCD; see the talks by H.\ Ohno and H.-T.\ Shu.

\subsection{Non-static screening masses\la{sec:nonstat}}

In this section we review some recent progress in the physics
interpretation of screening masses, especially the non-static ones.
Consider perturbing the QCD Hamiltonian $\hat H$ with an external
field $\phi$ coupling to a local QCD operator $\hat J$,
\be
\hat H_\phi(t) = \hat H - \int d^3y\;\phi(t,\vec y) \hat J(t,\vec y),
\qquad \qquad \phi(t,\vec y) = \delta(\vec y) e^{\omega t} \theta(-t), \qquad \omega\geq 0.
\ee
The latter equation describes a perturbation localized at the origin,
starting at $t=-\infty$ and shut off at $t=0$.  Let $G^{JJ}_E(\omega_n,\vec x)$ be the
Euclidean correlator, taken as a function of Matsubara frequency
$\omega_n$ and spatial separation $\vec x$. Linear response theory
leads to the following prediction for the change in the expectation
value of the operator $\hat J$ due to the external perturbation 
(see e.g.\ \cite{Meyer:2011gj},
\be\la{eq:GEJJ}
\delta \Big\<J(t=0,\vec x)\Big\>   =  {G^{JJ}_E(\omega_n,\vec x)},
 \qquad {\rm for~}\omega = \omega_n = 2\pi Tn.
\ee
This equation provides the physics interpretation of the static ($\omega_n=0$) and 
non-static Euclidean correlators. 
In particular, the correlation length in Matsubara sector $\omega_n$ is the
length scale over which a perturbation with the time dependence $
e^{\omega_n t}$ is screened ($n\geq 0$).
Eq.\ (\ref{eq:GEJJ}) also shows that the dependence of the correlation length
on $\omega$ can naturally be continued to continuous values of $\omega$ -- 
even though it is not directly accessible in the Matsubara formalism.

It is well-known that linear response, along with Fick's law $\vec J =
-D\nabla J_0$ as a constitutive equation for the vector current $\vec
J$, implies the existence of a diffusion pole in the retarded
correlator of the charge density $J_0$~\cite{Meyer:2011gj}.
Let $E(\omega_n)$ be the inverse correlation length of $J_0$ in Matsubara sector $\omega_n$.
In position space, one obtains
\be
G^{J_0J_0}_E(\omega_n,\vec x) \stackrel{\omega_n\to0, \vec x\to\infty}{=}
 -\chi_s\, E(\omega_n)^2~\; \frac{e^{-E(\omega_n) |\vec x|}}{4\pi |\vec x|}
\qquad{\rm with}\qquad 
E(\omega_n)^2 \stackrel{\omega_n\to0}{\sim} \frac{{\omega_n}}{{D}}.
\ee
Thus the hallmark of a diffusion pole is that an external
perturbation with the time dependence $e^{\omega t}$ coupling to $J_0$
is screened over a distance which diverges as $\sqrt{D/\omega}$ when
$\omega\to0$. 

Assuming that the diffusion pole exists and is continuously connected
to the $n=1$ Matsubara sector, the observation above suggests a new
method to extract the diffusion coefficient\footnote{Since the shear channel also contains a diffusion pole, the same
argument applies, with $D$ replaced by $\eta/(e+p)$, $\eta$ being the
shear viscosity and $e+p$ the enthalpy density.}.  If the correlation
length can be determined as a function of the Matsubara frequency and
continued to small frequencies, then $1/D$ can be
obtained as the slope of the graph of $E(\omega_n)^2$ at $\omega_n=0$.
An advantage of this approach is that at very high temperatures, where
$E(\omega_n)\simeq \omega_n$ for $\omega_n\geq 2\pi T$, the correct
result $1/D=0$ is obtained; this is in contrast with the method based
on the spectral function, where an extremely narrow transport peak
would have to be resolved.

This exercise has been carried out~\cite{Brandt:2014cka} in the
strongly coupled ${\cal N}=4$ SYM theory and shown to reproduce the
known result $D=(2\pi T)^{-1}$~\cite{Kovtun:2005ev}.  In QCD at weak
coupling, the relevant screening masses have been computed to ${\rm
  O}(g^2)$~\cite{Brandt:2014uda}; see the right panel of
Fig.\ \ref{fig:isosv}. This calculation is limited to the regime
where $\omega_n$ is much larger than the Debye mass.  On the
other hand, the weak coupling result~\cite{Arnold:2001ms} for the
diffusion coefficient is also displayed. A turnover of the
curve representing the screening masses must occur if it is to
make contact with kinetic theory.


Screening masses (static and non-static) in the vector channel have
been calculated recently in the high-temperature phase of $N_{\rm f}=2$ QCD
on fine lattices~\cite{Brandt:2014uda}.  The ${\rm O}(g^2)$
predictions for the non-singlet static screening masses at weak
coupling were made in~\cite{Laine:2003bd}, and these calculations were
extended to the non-static sectors in~\cite{Brandt:2014uda}. The basic
physics of these states is that they form non-relativistic bound
states of size ${\rm O}(m_E^{-1})$ from the point of view of a 2+1 dimensional
effective theory, where $m_E$ is the Debye screening mass.  Fairly  
good agreement is found between the weak-coupling predictions for the
screening masses and the lattice results in the non-static Matsubara
sector $n=1$.  An interesting connection between the weak-coupling
calculation of the screening masses and the dilepton production
rate~\cite{Aurenche:2002wq} was discovered~\cite{Brandt:2014uda},
which led to the realization that the same potential entering the
latter~\cite{CaronHuot:2008ni}, which was computed non-perturbatively
in~\cite{Panero:2013pla}, also determines the spectrum of non-static
screening masses. Using this non-perturbatively computed potential
improved the agreement with the lattice QCD results. The residues of
the screening states in the vector correlation function, which are
determined by the wave function of the non-relativistic bound state at
the origin, were however found to be significantly larger than the
weak-coupling calculations indicates, showing that the states are more
tightly bound than at weak-coupling.

\subsection{Numerical studies of spectral functions}

In the following, several channels are discussed where progress has
recently been made.  I address the channels in what I believe is an
increasing order of difficulty. We start with the case of the pion in
the low-temperature phase, where chiral effective theory provides
theory guidance. Furthermore, the pion parametrically dominates the
Euclidean correlators of the axial charge density
$\bar\psi\gamma_0\gamma_5\tau^a \psi$ and of the pseudoscalar density
$\bar\psi \gamma_5 \tau^a\psi$.

\noindent\emph{The pion quasiparticle in the low-$T$ phase.---}
Consider QCD in the SU(2) chiral limit at temperatures $T<T_c$; recall
that there must be a sharply defined phase transition, since the
chiral condensate $\<\bar\psi\psi\>$ is an order parameter.  The
spontaneous breaking of chiral symmetry implies the presence of a
divergent spatial correlation length (Goldstone theorem). Somewhat less 
obvious is the fact that a massless real-time
excitation exists, the pion quasiparticle. At small quark masses $m$, 
the inverse spatial correlation length (or `pion screening mass' $m_\pi$)
is of order $m_\pi \sim \sqrt{m}$. The low-momentum dispersion relation 
of the pion quasiparticle then has the form~\cite{Son:2002ci}
\be
\omega_{\vec p} = u (m_\pi^2+\vec p^2)^{1/2},
\ee
with $u$ a temperature-dependent parameter; it is unity at $T=0$ by
Lorentz symmetry. In the chiral limit, $u$ corresponds to the group
velocity of the pion. Thus a single pa\-ra\-meter
determines both the quasiparticle mass/screening mass ratio
$\omega_{\vec 0}/m_\pi$ and the $|{\vec p}|$ dependence of the
quasiparticle energy.

Exploratory calculations~\cite{Brandt:2014qqa} and a more accurate
calculation~\cite{Brandt:2015sxa} in two-flavor QCD with a
zero-temperature pion mass of 270MeV (at $T=170$MeV, which is below
$T_c$) show that $u\approx 0.74$ is significantly smaller than unity.
Also, the pion quasiparticle is found to be about 16\% lighter than
the zero-temperature pion.  This finding contradicts the picture of
the hadron resonance gas model, in which the excitations of the medium
are taken to be the same as at $T=0$. Recently, evidence was found
that the spectral density in the negative-parity nucleon sector also
exhibits a significant change below $T_c$~\cite{Aarts:2015mma}.
However, these findings are not necessarily inconsistent with the HRG
model providing successful predictions for equilibrium
quantities~\cite{Brandt:2015sxa}.

\noindent\emph{The heavy-quark diffusion constant.---}
Significant progress has also been made in the
study of the diffusion of a heavy quark in the quark-gluon plasma.
Heavy-ion collision phenomenology shows that in spite of their heavy
mass, charm quarks are `dragged along' by the medium, so that
they too exhibit an azimuthally anisotropic distribution~\cite{Kweon:2014kha}. 
The `kicks' of the medium on the heavy quarks therefore have to be strong enough
for it to equilibrate.

More precisely, in the static quark limit, the `kicks' are determined
by the color-electric force acting on the worldline of the heavy
quark~\cite{CaronHuot:2009uh}. The autocorrelation of the force
ultimately determines the diffusion constant $D$ of the heavy quark via an
Einstein relation,
\ba
G(x_0) &=& 
\frac{\Big\<\re\tr \big( U(\beta,x_0)gE_k(x_0,\vec{0})
                   U(x_0,0)gE_k(0,\vec{0})\big)\Big\>}
{ -3\; \< \re\tr U(\beta,0)\>} = \int_0^\infty {d\omega} \; \rho(\omega)\, 
K(x_0,\omega),
\nonumber\\
\kappa &=& \lim_{\omega\to0} {T} \rho(\omega)/\omega, \qquad D = 2T^2/\kappa.
\ea
Here $U(t_{\rm f},t_{\rm i})$ is the static quark propagator and $\vec E$ the color-electric field.
The spectral function has been calculated at next-to-leading order 
in perturbation theory~\cite{Burnier:2010rp}. It is found to be a smooth function at all frequencies.
This represents an advantage for lattice calculations, since resolving narrow features
in the spectral function is difficult. On the other hand, the coefficient $\kappa$
has been calculated at NLO~\cite{CaronHuot:2007gq}, 
and exhibits poor convergence at realistic values of the strong coupling.

Since the early lattice calculations~\cite{Meyer:2010tt}, 
continuum-extrapolated data have been analyzed in quenched
QCD~\cite{Kaczmarek:2014jga,Francis:2015daa}. As compared to the NLO
calculation, additional spectral weight is required at low frequencies
to describe the data. The analysis yields $1.8\leq\kappa/T^3\leq3.4$.
Although phenomenologically, radiative energy loss must be taken into
account too, this range is in the right ballpark to explain the
spectrum and anisotropic flow of heavy quarks.

\noindent\emph{The vector channel and the dilepton production rate.---}
The rate of production of dilepton pairs via virtual photons from a
thermal medium is directly proportional to the spectral function
$\rho^\mu{}_\mu$ of the electromagnetic current $J_\mu$ (see
e.g.\ \cite{Moore:2006qn} for a derivation).  Since the medium
produced in heavy-ion collisions appears to be in approximate local
thermal equilibrium, a weighted average of the spectrum of dileptons
over a range of temperatures is obtained; see for instance
\cite{Shen:2014vra}.

We will focus here on correlators of the spatial, non-singlet current $J_i$ at
vanishing spatial momentum, although preliminary results at
non-vanishing spatial momentum on fine, quenched ensembles have been
presented by Fl.\ Meyer at this conference.  At zero-temperature, the
spectral function $\rho_{ii}$ is dominated by the $\rho$ meson below
$\omega=1$GeV.  The main effect found in recent
calculations~\cite{Brandt:2012jc,Aarts:2014nba} is a shift of the
spectral weight from the $\rho$-meson region to the low-frequency
region $\omega\lesssim T$ as the temperature increases. The lattice
data in the high-temperature phase is consistent with a spectral
function containing the sum of a fairly broad transport
peak~\cite{Brandt:2012jc,Aarts:2014nba,Ding:2010ga} and the cut
contribution $\propto \omega^2$; such a spectral function lies
somewhere between the two thermal scenarios depicted in
Fig.\ \ref{fig:isosv}.  Estimates of the electric conductivity and the
diffusion coefficient $D$ have been presented. An important limitation
of calculations in dynamical QCD is the lack of a continuum limit of
the Euclidean correlator; the continuum limit has been taken in
quenched QCD~\cite{Ding:2010ga}.  Even setting this issue aside,
uncertainty remains on whether the (infinite-volume) spectral function
really is as smooth as the published results suggest.

A scenario where the methods used so far would yield completely
incorrect results for the electrical conductivity is if a small
fraction of the degrees of freedom contributing to the static
susceptibility had a very long mean-free-path; this would be extremely
hard to detect in the Euclidean correlator and would lead to a much
larger conductivity than the commonly used methods suggest. At the
same time, in a relatively fast hydrodynamic evolution of the system,
such degrees of freedom would then simply be frozen and not thermalize
locally with the rest of the medium.  Thus the large diffusion
coefficient in this scenario would also be less relevant. It seems
implausible to me that such weakly coupled degrees of freedom
exist in the high-$T$ phase of QCD.

Kinetic theory, which assumes the existence of weakly interacting
quasiparticles, predicts the presence of a transport peak in the
spectral function, with an area $\int_0^\Lambda
\frac{d\omega}{\pi\omega} \rho_{ii}(\omega)$ equal to $\chi_s\<v^2\>$,
where $\<v^2\>$ is the mean-square velocity of the
quasiparticles~\cite{Teaney:2006nc}\footnote{The cutoff $\Lambda$ is a separation scale
  between the inverse mean-free-time and the thermal scale.}. If the
quarks and gluons are the relevant quasiparticles at some temperature
above $T_c$, $\<v^2\>\approx1$ and the spectral weight of the transport
peak is known to a reasonable approximation. Thus if a transport peak
is resolved and has a spectral weight in the expected range, one can
be quite confident in the result. The main challenge at
high temperatures is thus to convincingly resolve the transport
peak of quark and gluons. 

The thermal modification $\Delta\rho(\omega,\vec k,T) \equiv \rho(\omega,\vec k,T)-\rho(\omega,\vec k,0)$
of the spectral functions in the vector and axial-vector spectral functions
is constrained by sum rules.
If we define the longitudinal and transverse spectral functions of the isovector
vector $V_\mu^a = \bar\psi \gamma_\mu \frac{\tau^a}{2}\psi$ and axial-vector 
$A_\mu^a = \bar\psi \gamma_\mu \gamma_5\frac{\tau^a}{2}\psi$      currents via
\ba
\frac{1}{3}\int d^3x \;e^{-i\vec k\cdot \vec x}\; \<V^a_0(x) V^a_0(0)\> &=& \int_0^\infty d\omega \,\rho_V^L(\omega,\vec k,T)\, 
K(x_0,\omega),
\\
\!\!\!\!\! - {\txts\frac{1}{6}}\Big(\delta_{il} - {k_ik_l}/{\vec k^2}\Big)
\int d^3x \;e^{-i\vec k\cdot \vec x}\; \<V^a_i(x) V^a_l(0)\> &=& \int_0^\infty d\omega \,\rho_V^T(\omega,\vec k,T)\, 
K(x_0,\omega), \qquad 
\\
\frac{1}{3}\int d^3x\; e^{-i\vec k\cdot\vec x}\;\< A_0^a(x) A_0^a(0)\> 
&=& \int_0^\infty d\omega\, \rho_{_{\rm A}}^L(\omega,\vec k,T)\,
K(x_0,\omega) \qquad \quad 
\ea
we have the sum rules~\cite{Bernecker:2011gh,Brandt:2014qqa},
\ba\la{eq:SR1}
\frac{1}{2\pi}\int_{-\infty}^\infty {d\omega}\,{\omega}\, \Delta\rho_V^{L}(\omega,\vec k,T) &=& 0, \qquad\qquad\qquad 
\forall\vec k 
\\  \la{eq:SR3}
\frac{1}{2\pi} \int_{-\infty}^\infty \frac{d\omega}{\omega} \Delta\rho_V^{L}(\omega,\vec k,T) &=& \chi_s - \kappa_l {\vec k}^{\,2} + {\rm O}(|\vec k|^4),
\\ \la{eq:SR2}
\frac{1}{2\pi}\int_{-\infty}^\infty \frac{d\omega}{\omega} \Delta\rho_V^T(\omega,\vec k,T) &=& \kappa_t {\vec k}^{\,2} + {\rm O}(|\vec k|^4),
\\ \la{eq:SR4}
\frac{1}{2\pi}\int_{-\infty}^\infty {d\omega\; \omega} \;\Delta\rho_{_{\rm A}}^L(\omega,\vec k,T)
&=&  -m\<\bar\psi\psi\>\Big|^{T}_{0},\qquad \qquad \forall \vec k. 
\ea
A physics interpretation of $\kappa_l$ and $\kappa_t$ in terms of
screening/antiscreening of electric probe charges and currents placed
in the medium exists~\cite{Brandt:2013faa}.  These sum rules have been
used to constrain the thermal change in the spectral functions. 
 Note that at vanishing spatial momentum, $\rho^T=\frac{1}{3} \rho_{ii}$ and Eq.\ (\ref{eq:SR2}) simplies to
 $\int_0^\infty \frac{d\omega}{\omega} \Delta\rho_{ii}(\omega,T)=0$.

\noindent\emph{Quarkonium in the high-temperature phase.---}
The physics of quarkonium in the high-temperature phase has a long
history in the heavy-ion community~\cite{Matsui:1986dk}.  At what
temperature different $\bar cc$ and $\bar bb$ quasiparticles melt
provides a `thermometer' in heavy-ion collisions: $p$-wave bound
states are expected to `melt' at a lower temperature than $s$-wave
bound states.  A sequential suppression of the $\Upsilon$ states has
been observed in dimuon spectra by
CMS~\cite{Chatrchyan:2011pe}, providing an interval 
for the maximum temperature reached in Pb+Pb collisions.

The mechanism of the disappearance of $\bar QQ$ quasiparticles has
been investigated in detail at weak coupling
(see~\cite{Laine:2008cf} and Refs. therein).   
The Debye screening of the $\bar QQ$ potential and
the scattering with the partons of the medium induce a decrease 
of spectral weight at treshold in the spectral function of $\bar Q\, \vec\gamma Q$
(see Fig.~2 of~\cite{Laine:2007gj}).

Charmonium quasiparticles have mostly been studied using the full
relativistic formulation (e.g.\ in quenched QCD~\cite{Ding:2012sp}, and in
$N_{\rm f}=2+1$ QCD with heavy Wilson fermions\cite{Borsanyi:2014vka}).
S-wave charmonium quasiparticles show up in the spectral function up
to about $1.4T_c$.  We note that the sum rule (\ref{eq:SR2}) applies
in the $J/\psi$ channel and could be used to constrain the thermal
changes to the spectral function.  

S.\ Kim presented a study in non-relativistic QCD (NRQCD) at this
conference.  The pros of using NRQCD are that the exponential kernel
$G(x_0)= \int_{-2m_Q}^\infty \frac{d\omega}{2\pi}\;e^{-\omega
  x_0}\;\rho(\omega)$ allows for a better $\omega$-resolution; that
$\rho(\omega)$ is softer in the ultraviolet than in the relativistic
theory; that $\rho(\omega)$ does not contain a transport peak; and
that very accurate data can be obtained. All this makes the NRQCD
correlator more sensitive to changes in the spectral weight near the
$\bar QQ$ threshold.  On the downside, dealing with cutoff effects is
more delicate in NRQCD.

Bottomomium quasiparticles, due to their large mass, have been
addressed using the effective field theories NRQCD and potential NRQCD
(pNRQCD). A recent NRQCD calculation~\cite{Aarts:2014cda} in 2+1
flavor QCD with $m_\pi=400$MeV finds that while the ground state
$\Upsilon$ ($J^{PC}=0^{-+}$) survives at least up to $2T_c$, the
$\chi_{b_1}$ ($J^{PC}=1^{++}$) melts immediately above $T_c$.
Ref.\ \cite{Kim:2014iga} on the other hand finds that the $\chi_{b_1}$
survives for some interval of temperature above $T_c$. Part of the
difficulty is that the spectral weight near threshold does not seem to
be the dominant one. It would be interesting for future studies to
quote the integral of the spectral function over the threshold region,
since it is less sensitive to the method used
for the inverse problem. Finally, we note that a pNRQCD calculation on
the 2+1 flavor ensembles of the HotQCD collaboration has recently
emerged~\cite{Burnier:2015tda}.



\noindent\emph{The nucleon channel.---}
While many bosonic channels have been investigated, the fermionic
channels have received very little attention until
recently~\cite{Aarts:2015mma,Datta:2012fz}.  One motivation for
looking at baryon correlation functions is that they are a probe of
chiral symmetry restoration. To be specific, let $O_{N+} =
\frac{1}{2}(1+\gamma_0) \,{\epsilon_{abc}(u_a C\gamma_5 d_b) u_c}$ be
a nucleon interpolating operator (it has an intrinsic positive
parity).  Consider then its two-point function and corresponding
spectral representation~\cite{Aarts:2015mma},
\be
G(x_0) = \int d^3x\; \< O_{N+}(x) \;\bar O_{N+}(0)\> 
=\int_{0}^\infty \frac{d\omega}{2\pi}\; \frac{1}{1+e^{\omega/T}}\;   
 \Big[ \rho_+(\omega) e^{-\omega x_0} - \rho_-(\omega) e^{-\omega(\beta-x_0)}\Big].
\ee
The restoration of chiral symmetry implies parity doubling: $G(\beta-x_0) = G(x_0)$.
Therefore the ratio 
$ R(x_0) =  \frac{G(x_0)-G(\beta-x_0)}{G(x_0)+G(\beta-x_0)}$
is an order parameter for chiral symmetry restoration. The same
argument applies to static correlators~\cite{Datta:2012fz}, which
addresses the degeneracies between screening masses.

Both the static correlators (in a quenched study \cite{Datta:2012fz})
and the temporal correlators~\cite{Aarts:2015mma} exhibit the expected
behavior that the ratio $R$ starts to drop rapidly at $T=T_c$. 
Approximating the spectral functions $\rho_+$ and $\rho_-$ 
by a single Dirac delta-function $A_\pm \delta(\omega-m_\pm)$
also show that the relative mass difference $(m_- -m_+)/(m_- + m_+)$ drops as $T_c$ is
approached from below~\cite{Aarts:2015mma}.
While little thermal change was found in the positive-parity sector, 
in principle even the nucleon acquires a thermal width through resonant collisions 
with the pions of the medium, $N\pi \longleftrightarrow \Delta$.

How strongly baryon quasiparticles are modified below $T_c$, and
whether any of them survive for a short interval of temperature above
$T_c$ would be as interesting to know as for the mesons. It may also
help understand what happens when a baryon chemical potential is
turned on. 
 \vspace{-0.1cm}

\section{Conclusion\la{sec:concl}}
  \vspace{-0.1cm}

A lot has been learnt about thermal QCD from the lattice. It turns out
that the crossover nature of the QCD thermal transition is a very
robust feature with respect to varying the quark masses. Efforts to
determine the nature of the transition at very small $m_{ud}$ are
ongoing. For definitive results, I believe simulating at several 
lattice spacings $a\lesssim 0.08{\rm fm}$ will be required.
For a lattice action with exact chiral symmetry, the requirements can perhaps
be relaxed somewhat.

There has been significant progress in the study of near-equilibrium
properties via spectral functions.  Lattices with $N_t\approx24$
points in the Euclidean time direction have become available, the
continuum limit has been taken in quenched QCD and the statistical
precision in several channels is now at the permille level.  
Statistical errors have decreased by an order of magnitude since 2007.

There were many interesting talks about finite-temperature, and
indeed about finite-density QCD at this conference, demonstrating the
vitality of the field.  I thank the organizers for the invitation and
for running a very pleasant and informative conference.  My
own research is supported by the DFG grant ME 3622/2-1
\emph{Static and dynamic properties of QCD at finite temperature.}

\bibliographystyle{JHEP}
\bibliography{/Users/harvey/BIBLIO/viscobib}


\end{document}